\documentstyle[a4,12pt]{article}
\begin{document}
\pagestyle{empty}
\begin{titlepage}
\rightline{UTS-DFT-95-07}
\rightline{IC/95/293}
\vspace{1.0 truecm}
\begin{center}
\begin{Large}
{\bf Role of beam polarization in the determination of $WW\gamma$ and
$WWZ$ couplings from $e^+e^-\to W^+W^-$}
\end{Large}

\vspace{1.5cm}

{\large  V.V. Andreev\hskip 2pt\footnote{Permanent address: Gomel
State University, Gomel, 246699 Belarus.}
}\\[0.3cm]
International Centre for Theoretical Physics, Trieste, Italy

\vspace{5mm}

{\large  A.A. Pankov\hskip 2pt\footnote{Permanent address: Gomel
Polytechnical Institute, Gomel, 246746 Belarus.\\
E-mail PANKOV@GPI.GOMEL.BY}
}\\[0.3cm]
International Centre for Theoretical Physics, Trieste, Italy\\
Istituto Nazionale di Fisica Nucleare, Sezione di Trieste, 34127 Trieste,
Italy

\vspace{5mm}

{\large  N. Paver\hskip 2pt\footnote{Also supported by the Italian
Ministry of University, Scientific Research and Technology (MURST).}
}\\[0.3cm]
Dipartimento di Fisica Teorica, Universit\`{a} di Trieste, 34100
Trieste, Italy\\
Istituto Nazionale di Fisica Nucleare, Sezione di Trieste, 34127 Trieste,
Italy\\
\end{center}

\vspace{1.cm}

\begin{abstract}
\noindent
We evaluate the constraints on anomalous trilinear gauge-boson couplings
that can be obtained from the study of electron-positron annihilation into
$W$ pairs at a facility with either the electron beam longitudinally polarized
or both electron and positron beams transversely polarized. The energy ranges
considered in the analysis are the ones relevant to the next-linear collider
and to LEP~200. We discuss the possibilities of a model independent analysis
of the general $CP$ conserving anomalous effective Lagrangian, as well as its
restriction to some specific models with reduced number of independent
couplings. The combination of observables with initial and final state
polarizations allows to separately constrain the different couplings and to
improve the corresponding numerical bounds.\par

\vspace*{3.0mm}

\end{abstract}
\end{titlepage}

\pagestyle{plain}
\setlength{\baselineskip}{1.3\baselineskip}
\section{Introduction}
The experimental confirmation of the Standard Model (SM) is presently
limited to the sector of the interaction of fermions with vector bosons
\cite{langacker1}, where an impressive agreement is found. Another key
ingredient of the SM, not tested yet, is the
interaction in the gauge-boson sector, which follows from the non-abelian
structure of the electroweak symmetry and assures the renormalizability of
the theory. Accordingly, in the physics programme at planned high energy
(and high luminosity) colliders, much emphasis is given to precise
measurements of the $WW\gamma$ and $WWZ$ couplings. Such measurements should
eventually confirm the SM, or maybe discover `anomalous' values of these
couplings indicating physics beyond the SM.\par
While experiments at low energy and precision measurements in $e^+e^-$
annihilation at the $Z^0$ pole can provide {\it indirect} access to these
constants \cite{zeppenfeld1,hernandes,ruj}, only high energy colliders,
well above the threshold for $W$-pair production, will allow {\it direct}
and unambiguous tests. In this regard, some limits are available from Tevatron
\cite{tevatron}, and in the near future one can foresee experimental studies
of boson self-couplings at LEP~200 \cite{hagiwara1}-\cite{sekulin},
and to some extent at HERA \cite{hera}. A new stage in precision will be
reached at the planned hadron collider
LHC \cite{lhc,baur} and at the $e^+e^-$ linear colliders \cite{nlc},
taking advantage of the increased sensitivity to deviations from the SM
allowed by the significantly higher CM energies of these machines.\par
A particularly sensitive process where to study the trilinear gauge boson
couplings is the $W$-pair production \cite{hagiwara1}
\begin{equation}e^++e^-\to W^++W^-.\label{proc}\end{equation}
In this process the enhanced sensitivity to anomalous values of those
couplings reflects the partial compensation among the individual,
$\sqrt s$-diverging contributions to the SM cross section ($\sqrt s$ is the
CM energy), corresponding at the Born level to $\gamma$, $\nu$ and $Z$
exchange diagrams and their interferences. Instead, the SM couplings are such
that the gauge cancellation exactly occurs in the asymptotic regime, and
consequently the SM cross section has a
decreasing behavior with $\sqrt s$ \cite{alles}.\par
Considering the following modification of the
$\gamma$- and $Z$-exchange amplitudes ($V=\gamma,\hskip 2pt Z$):
\begin{equation}
{\cal A}(V)\to{\cal A}(V)+\Delta{\cal A}(V)=(1+f_V){\cal A}(V),
\label{aplit}\end{equation}
where $f_V$'s linearly depend on the anomalous couplings, and introducing
the relative deviation from the SM prediction for the cross
section (either total, or differential, or integrated in some angular range):
\begin{equation}\Delta\equiv\frac{\Delta\sigma}{\sigma_{SM}}=
\frac{\sigma_{anom}-\sigma_{SM}}{\sigma_{SM}},
\label{delta}\end{equation}
one has:
\begin{eqnarray}
\sigma_{anom}&\propto&\vert{\cal A}_1(\nu)+(1+f_\gamma){\cal A}(\gamma)
+(1+f_Z){\cal A}(Z)\vert^2+\vert {\cal A}_{2}(\nu)\vert^2,
\nonumber \\
\sigma_{SM}&\propto&\vert {\cal A}_1(\nu)+{\cal A}(\gamma)+
{\cal A}(Z)\vert^2+\vert {\cal A}_2(\nu)\vert^2.
\label{sig}\end{eqnarray}
Here, taking into account the $W^-$ ($W^+$) helicities $\tau$ ($\tau^\prime$),
for later convenience the neutrino-exchange amplitude has been divided into
the $\vert\tau-\tau^\prime\vert\leq1$ part ${\cal A}_1(\nu)$ plus the
$\vert\tau-\tau^\prime\vert=2$ part ${\cal A}_2(\nu)$. As evident in
Eq.~(\ref{sig}), the amplitude ${\cal A}_2(\nu)$ does not interfere with the
others. To first order in $f_V$ one obtains:
\begin{equation}\Delta=\Delta_{\gamma}+\Delta_Z,\label{del}\end{equation}
where
\begin{eqnarray}\Delta_\gamma=f_{\gamma}(R_{\nu\gamma}+R_{Z\gamma}
+2R_{\gamma\gamma}),
\nonumber \\
\Delta_Z=f_Z(R_{\gamma Z}+R_{\nu Z}+2R_{ZZ}),
\label{delgz}\end{eqnarray}
and ($i,j=\gamma,\nu,Z$)
\begin{equation}R_{ij}=\sigma_{ij}/\sigma_{SM};\qquad
\sigma_{SM}\equiv\sigma(f_V=0)=\sum_{i,j}
\sigma_{ij}.\label{rij}
\end{equation}
In Eqs.~(\ref{del}-\ref{rij}), $\Delta$'s are determined by
linear combinations of non
cancelling individually divergent contributions, and will increase, basically
like a power of $s$. In contrast, the SM cross section decreases at least
as $1/s$. Thus, if we parametrize the sensitivity of process (\ref{proc}) to
$f_V$ by, e.g., the ratio
${\cal S}={\Delta/(\delta\sigma/\sigma)}$, with
${\delta\sigma/\sigma}$ the statistical uncertainty experimentally attainable
on the SM cross section, such a sensitivity is power-like enhanced with
increasing $\sqrt s$, even at fixed integrated luminosity, namely
${\cal S}\propto\sqrt{ L_{int}\cdot s}$.\footnote{This behavior of
the sensitivity of process (\ref{proc}) also applies to other non-standard
effects, such as, e.g., $Z-Z^\prime$ mixing \cite{pankov1} and lepton mixing
\cite{pankov2}.}\par
As discussed in Ref.~\cite{pankov1}, a dramatic improvement in the sensitivity
to anomalous values of $WW\gamma$ and $WWZ$ vertices should be obtained if
the initial electron beam were longitudinally polarized, and one could
separately measure the cross sections for both $e^-_Le^+$ ($\sigma^{L}$) and
$e^-_Re^+$ ($\sigma^{R}$) annihilation. In particular, although being
suppressed by $\gamma-Z$ compensation and thus leading to lower statistics,
$\sigma^{R}$ has the advantage of being free of the neutrino-exchange
contribution. In general, ${\cal A}_2(\nu)$ numerically dominates the SM
$\sigma^{unpol}$ and $\sigma^L$, and is not modified by anomalous trilinear
couplings, so that it tends to diminish the sensitivity of these cross
sections to such effects. Consequently, one can qualitatively expect
$\sigma^R$ to allow improved constraints even in the
case of just one anomalous coupling taken as a free parameter.\par
In addition, for two (or more) free parameters, by themselves the cross
sections $\sigma^L$ and $\sigma^R$ separately provide correlations among
parameters rather than limits. In fact, from Eq.~(\ref{sig}) and the
approximate relation ${\cal A}(\gamma)\approx
-{\cal A}^R(Z)\approx{\cal A}^L(Z)$ at $\sqrt s\gg M_Z$, the deviations of
$\sigma^L$ and $\sigma^R$ from the SM are easily seen to bring information on
the following combinations:
\begin{eqnarray}\Delta\sigma^L\propto f_{\gamma}+f_Z,
\nonumber \\
\Delta\sigma^R\propto f_{\gamma}-f_Z.\label{correl}\end{eqnarray}
Due to $\sigma^L\gg\sigma^R$, also $\sigma^{unpol}\propto f_{\gamma}+f_Z$.
Clearly, the combination of $\sigma^R$ and $\sigma^L$ (or $\sigma^{unpol}$)
could be essential in order to significantly reduce the allowed region in the
$(f_\gamma,\ f_Z)$ plane, by the intersection of the `orthogonal' correlation
areas provided by Eq.~(\ref{correl}).\par
For realistic values of the electron longitudinal polarization, less
than 100\%, the determination of $\sigma^R$ from the data could be
contaminated by the uncertainty in the polarization itself, which allows the
presence of some left-handed cross section. Due to
${\sigma^{L}\gg\sigma^{R}}$, such an uncertainty could induce a systematic
error on $\sigma^{R}$ larger than the statistical error for this cross
section, and consequently the sensitivity would be diminished. However, as
shown in Ref.~\cite{pankov1}, one can find `optimal' kinematical cuts
to drastically reduce this effect.\par
In the general $CP$ conserving case, the anomalous effective Lagrangian for
trilinear gauge boson couplings depends on five constants, which are
difficult to disentangle from each other by using just the unpolarized cross
section, not only due to the large number of parameters, but also due to
possible accidental cancellations which might reduce the sensitivity of this
observable. To separate the coupling constants, and constrain their values in
a model independent way, measurements of the cross sections for polarized
final $W$'s and both initial longitudinal polarizations should be combined.
\par
In this paper we will present an estimate, along the lines exposed above, of
the bounds on the anomalous three-boson coupling constants that can be
obtained from the analysis of the process  $e^+e^-\to W^+W^-$ based on the
combination of polarized cross sections, at the reference energy of the
planned $e^+e^-$ linear colliders, namely $0.5$ up to $1\hskip 2pt TeV$, with
polarized electron beams and assuming that also $W^+W^-$ polarization will be
measured. A general discussion of the prospects and feasibility of measuring
polarization effects in $W$-pair production can be found, e.g., in
Refs.~\cite{bilenky1} and \cite{blondel,frank}.\par
Also, we study the region around $\sqrt s=200\ GeV$ appropriate to LEP~200,
since this machine will be operational in a relatively near future. Here,
initial longitudinal
polarization will not be available so that only unpolarized or transverse
beam polarization will exist. The latter is an attractive option at $e^+e^-$
storage rings like LEP~200 \cite{zeppenfeld3,fleischer}, where electron and
positron spins naturally align in opposite directions in the magnetic field of
the accelerator. The transverse polarization could be exploited to perform the
model independent analysis, assuming the possibility of measuring
final $W^+W^-$ polarizations also in this case. In fact, for the
`transverse' azimuthal asymmetry $A_T$ (precisely defined in the sequel),
one has $A_T\propto {\cal A}^L\hskip 2pt {\cal A}^R$  with
${\cal A}^L={\cal A}(\nu)+{\cal A}(\gamma)+{\cal A}^L(Z)$ and
${\cal A}^R={\cal A}(\gamma)+{\cal A}^R(Z)$ and, due to
${\cal A}^L\gg{\cal A}^R$, the deviation from the SM model is
$\Delta A_T\propto f_{\gamma}-f_Z$. Consequently, to obtain the allowed
region in the $(f_\gamma,\ f_Z)$ plane, in this case $A_T$ plays the same
role as $\sigma^R$ in the previous example in Eq.~(\ref{correl}).\par
Specifically, in Sect.~2 we will introduce the standard parameterization of
the $WW\gamma$ and $WWZ$ vertices and will briefly review current and expected
bounds on these parameters from forthcoming experiments. In Sect.~3 we
introduce the helicity amplitudes and the corresponding observables relevant
to our analysis and in Sect.~4 we present the resulting constraints (model
independent as well as model dependent ones) on the anomalous couplings from
future $e^+e^-$ linear colliders. Sect.~5 is devoted to a similar analysis
at the energy of LEP~200 and, finally, Sect.~6 contains some concluding
remarks. Formulae relevant to the cross sections needed for our numerical
analysis are collected in an Appendix.

\section{Trilinear gauge-boson vertices}

We limit to the $C$ and $P$ invariant part of the $WWV$ interaction, which in
general can be represented by the effective Lagrangian with five independent
couplings \cite{gounaris1}:
\begin{eqnarray}{\cal L}_{eff}&=&-ie\hskip 2pt
\left[A_\mu\left(W^{-\mu\nu}W^+_\nu-
W^{+\mu\nu}W^-_\nu\right)+F_{\mu\nu}W^{+\mu}W^{-\nu}\right]-ie\hskip 2pt
x_{\gamma}\hskip 2pt F_{\mu\nu}W^{+\mu}W^{-\nu}
\nonumber\\
&-& ie\hskip 2pt \left(\cot\theta_W+\delta_Z\right)\hskip 2pt \left[Z_\mu
\left(W^{-\mu\nu}W^+_\nu-W^{+\mu\nu}W^-_\nu\right)+Z_{\mu\nu}W^{+\mu}W^{-\nu}
\right]\nonumber\\
&-&ie\hskip 2pt x_Z\hskip 2pt Z_{\mu\nu}W^{+\mu}W^{-\nu}
 +ie\hskip 2pt\frac{y_{\gamma}}{M_W^2}\hskip 2pt F^{\nu\lambda}W^-_{\lambda\mu}
W^{+\mu}_{\ \ \nu}+ie\hskip 2pt\frac{y_Z}{M_W^2}\hskip 2pt
Z^{\nu\lambda}
W^-_{\lambda\mu}W^{+\mu}_{\ \ \nu}\hskip 2pt ,\label{lagra}\end{eqnarray}
where $W_{\mu\nu}^{\pm}=\partial_{\mu}W_{\nu}^{\pm}-
\partial_{\nu}W_{\mu}^{\pm}$ and ${Z_{\mu\nu}=\partial_{\mu}Z_{\nu}-
\partial_{\nu}Z_{\mu}}$.
In Eq.~(\ref{lagra}), $e=\sqrt{4\pi\alpha_{em}}$ and $\theta_W$ is the
electroweak angle. The relation of the above constants to those more
directly connected with $W$ static properties is
\begin{eqnarray}
&x_\gamma \equiv \Delta k_\gamma=k_\gamma-1; \hskip 10pt
y_\gamma \equiv \lambda_\gamma,& \nonumber\\
\delta_Z \equiv g_Z-\cot\theta_W;\hskip 7pt &
x_Z \equiv \Delta k_Z\hskip 2pt (\cot\theta_W+\delta_Z)=(k_Z-1)\hskip 2pt
g_Z;\hskip 7pt &  y_Z \equiv \lambda_Z \cot\theta_W.\label{param}
\end{eqnarray}
With $\mu_W$ and $Q_W$ the $W$ magnetic and quadrupole electric moments,
respectively:
\begin{equation}
\mu_W={e\over {2M_W}}(1+k_{\gamma}+\lambda_{\gamma})\quad ,\qquad
Q_W=-{e\over {M_W^2}}(k_{\gamma}-\lambda_{\gamma}),
\end{equation}
and a similar interpretation holds for the $WWZ$ couplings.\par
At the tree-level, the SM values of these couplings are
\begin{equation}
\delta_Z=x_\gamma=x_Z=y_\gamma=y_Z=0.
\label{smparam}
\end{equation}
In the SM, the natural size of $\Delta k_\gamma$ and $\lambda_\gamma$ is
$\alpha_{em}/\pi\sim 10^{-3}$ \cite{bardeen}. In extensions of the SM
such as those containing extra Higgs doublets, extra heavy fermions
\cite{couture1}, or supersymmetric extensions \cite{couture2,lahanas},
the deviations from the tree-level SM values tend to be of the same
order of magnitude as these one-loop corrections.\par
Briefly summarizing the present information and the future perspectives
concerning the anomalous couplings, {\it indirect} constraints on $WW\gamma$
and $WWZ$ vertices have been obtained by comparing low energy data
($\sqrt s<2\hskip 2pt M_W$) with SM predictions for observables that can
involve such vertices at the loop level \cite{hernandes,burgess}. These
limits are derived from a global analysis of the data varying one parameter
at a time and keeping the remaining ones fixed at the SM values, and are
relatively weak with respect to the size of the SM corrections:
$\vert\Delta k_\gamma\vert\leq0.12$, $\vert\Delta k_Z\vert\leq0.08$,
$\vert\lambda_\gamma\vert\leq 0.07$, and $\vert\lambda_Z\vert\leq 0.09$ at
$95\%$ CL \cite{burgess}. \par
Direct tests of trilinear gauge boson couplings at higher energies
($\sqrt s>2\hskip 2pt M_W$) have been attempted in
$p\bar p\to W^\pm\gamma,\ W^\pm Z$, and $W^+W^-$ at the
Tevatron, still considering one constant at a time as a free parameter
\cite{tevatron}. In this case, limits are of the order of unity, and
therefore are not yet stringent enough to significantly test the SM. The
expected sensitivities from future Tevatron experiments are
$\vert\Delta k_\gamma\vert$,
$\vert\lambda_\gamma\vert\sim{\cal O}(0.1)$ at $\int Ldt=1\ fb^{-1}$, and
in the longer term the hadron collider LHC would improve the Tevatron bounds
for $\Delta k_{\gamma,Z}$ and $\lambda_{\gamma,Z}$ to an accuracy in the
range ${\cal O}(0.01-0.1)$, assuming an integrated luminosity of
$100\ fb^{-1}$ \cite{baur}.\par
In the near perspective, some constraint on the $WW\gamma$ vertex to an
accuracy of about $\pm 0.5$ should be obtainable at HERA from single $W$
production \cite{hera}. \par
Indeed, the test of the trilinear gauge boson couplings from the $W$ pair
production process (\ref{proc}) will be one of the major items in the
forthcoming physics programme at LEP~200 \cite{hagiwara1}-\cite{sekulin},
where an accuracy of ${\cal O}(0.1)$ is expected from direct measurements of
the cross section.\par
In the more distant future, the next linear $e^+e^-$ colliders (NLC), with
$\sqrt s\geq 500\hskip 2pt GeV$ \cite{gounaris1,aihara},
will probably provide the best opportunities to analyse gauge boson couplings
with significant accuracy from the $W^\pm$ pair production process
(\ref{proc}), due to the really high sensitivity of this reaction
at such energies, in particular if initial
beam polarization will be available. Depending on the CM energy and the
integrated luminosity, it should be possible to test those couplings {\it via}
a model independent analysis, and look for deviations from
the SM with an accuracy up to some units$\times 10^{-3}$. \par
In the next Section, we present the helicity amplitudes and the polarized
observables relevant to the analysis of process (\ref{proc}).

\section{Helicity amplitudes and polarized cross sections}
In Born approximation, process (\ref{proc}) is described by the $\nu$,
$\gamma$ and $Z$ exchange amplitudes in Fig.~1. The differential cross
section for initial $e^+_{\lambda^\prime}e^-_{\lambda}$ and final
$W^+_{\tau^\prime} W^-_{\tau}$ states can be expressed as
\begin{equation}
\frac{d\sigma^{\lambda\lambda^\prime}_{\tau\tau^\prime}}{d\cos\theta}=
\frac{\vert\vec p\vert}{4\pi s\sqrt{s}}
\vert {\cal A}_{\tau\tau^\prime}^{\lambda\lambda^\prime}(s,\ \cos\theta)
\vert^2.\label{difcros}\end{equation}
In Eq.~(\ref{difcros}), $\vert\vec p\vert=\beta_W\sqrt{s}/2$,
$\beta_W=\sqrt{1-4M_W^2/s}$, $\lambda=\pm 1/2$ with
$\lambda^\prime=-\lambda$ represents the electron (positron) helicities,
and $\tau\hskip 2pt (\tau^\prime)=\pm 1,\hskip 2pt 0$ are the
$W^-$ ($W^+$) helicities.\par
The helicity amplitudes ${\cal A}_{\tau\tau^\prime}^{\lambda\lambda^\prime}$
are listed in Tab.~\ref{tab:tab1}, in a form convenient to our analysis
\cite{gounaris1}.
Notations are such that $t=M_W^2-s(1-\beta_W\cos\theta)/2$,
$v=(T_{3,e}-2Q_e\hskip 2pt s_W^2)/2s_Wc_W$ and $a=T_{3,e}/2s_Wc_W$, where
$t$ is the momentum transfer and $v$ and $a$ are, respectively, the SM vector
and axial-vector couplings of electrons to the $Z$ boson
($s_W=\sin\theta_W$, $c_W=\cos\theta_W$). The first column in
Tab.~\ref{tab:tab1} contains the relevant combinations
of coupling constants and propagators, while the remaining two contain
kinematical factors. In order to obtain the amplitude for definite electron
helicity $\lambda$ and $W^\mp$ helicities $\tau$ and $\tau^\prime$, one has to
sum the products of all the relevant entries in the first column times the
corresponding kinematical factor in the same row times the common
kinematical factor on top of the second (or of the third) column. \par
In a circular storage ring collider, such as LEP~200, transverse polarization
of electron and positron beams can naturally occur. Thus, introducing for
$e^-$ and $e^+$ the magnitudes of longitudinal and transverse polarizations,
$P_L,P_L^\prime$ and $P_T,P_T^\prime$, the averaged square of the matrix
element for arbitrarily polarized initial beams can be written
as \cite{zeppenfeld3,fleischer}:
\begin{eqnarray}{\overline{\vert {\cal A}\vert^2}}&=&\frac{1}{4}
\{(1-P_LP_L^\prime)[\vert {\cal A}^+\vert^2+\vert{\cal A}^-\vert^2]
+(P_L-P_L^\prime)[\vert {\cal A}^+\vert^2-\vert {\cal A}^-\vert^2]
\nonumber \\
&+& 2P_TP_T^\prime[ \cos(2\phi_W){\rm Re} ({\cal A}^+{\cal A}^{-*})
-\sin(2\phi_W){\rm Im} ({\cal A}^+{\cal A}^{-*})]\},\label{nat}\end{eqnarray}
where $\phi_W$ is the azimuthal production angle of the $W^-$ and
${\cal A}^\pm$ correspond to $\lambda=-\lambda^\prime=\pm 1/2$ for arbitrary
$W^\mp$ helicities $\tau,\hskip 2pt\tau^\prime$. \par
\begin{table}
\centering
\caption{Helicity amplitudes for $e^+e^-\to W^+W^-$}
\begin{tabular}{|c|c|c|}
\hline
$e^+_{-\lambda} e^-_\lambda\to W^+_LW^-_L$  &
$ \tau=\tau^\prime=0$&$$\\
$$&$-\frac{e^2 S\lambda}{2}\sin\theta$ & $$\\ \hline
$\frac{2\lambda-1}{4\hskip 2pt t\hskip 2pt s^2_W}$ &
$\frac{S}{2M_W^2}[\cos\theta-\beta_W (1+\frac{2M_W^2}{S})]$ &\\
\hline
$-\frac{2}{S}+\frac{2\cot\theta_W}{S-M_Z^2}(v-2a\lambda)$ &
$-\beta_W(1+\frac{S}{2M^2_W})$ &\\ \hline
$ -\frac{x_\gamma}{S}+
\frac{x_Z+\delta_Z (3-\beta_W^2)/2}{S-M_Z^2}(v-2a\lambda)$ &
$-\beta_W\frac{S}{M^2_W}$ &\\ \hline
\hline
$e^+_{-\lambda} e^-_\lambda\to W^+_TW^-_T$ & $\tau=\tau^\prime=
\pm 1$ & $\tau=-\tau^\prime=\pm 1$ \\
& $-\frac{e^2 S\lambda}{2}\sin\theta$
& $-\frac{e^2 S\lambda}{2}\sin\theta$\\
\hline
$\frac{2\lambda-1}{4\hskip 2pt t\hskip 2pt s^2_W}$ &
$\cos\theta-\beta_W $ & $-\cos\theta-2\tau\lambda $\\
\hline
$-\frac{2}{S}+\frac{2\cot\theta_W}{S-M_Z^2}(v-2a\lambda)$ &
$-\beta_W $ & $0$ \\ \hline
$-\frac{y_\gamma}{S}+
\frac{y_Z+\delta_Z (1-\beta_W^2)/2}{S-M_Z^2}(v-2a\lambda)$ &
$-\beta_W\frac{S}{M^2_W}$ & $0$ \\ \hline
\hline
$e^+_{-\lambda} e^-_\lambda\to W^+_TW^-_L$ &
$\tau=0$, $\tau^\prime=\pm 1$ &
$\tau=\pm 1$, $\tau^\prime=0$ \\
& $-\frac{e^2 S\lambda}{2\sqrt{2}}(\tau^\prime\cos\theta-2\lambda)$
& $\frac{e^2 S\lambda}{2\sqrt{2}}(\tau\cos\theta+2\lambda)$\\
\hline
$\frac{2\lambda-1}{4\hskip 2pt t\hskip 2pt S^2_W}$ &
$\frac{\sqrt{S}}{2M_W}[\cos\theta(1+\beta_W^2)-2\beta_W]- $ &
$\frac{\sqrt{S}}{2M_W}[\cos\theta(1+\beta_W^2)-2\beta_W]-$\\
& $-\frac{2M_W}{\sqrt{S}}\frac{\tau^\prime\sin^2\theta}
{\tau^\prime\cos\theta-2\lambda}$ &
$-\frac{2M_W}{\sqrt{S}}\frac{\tau\sin^2\theta}
{\tau\cos\theta+2\lambda}$ \\
\hline
$-\frac{2}{S}+\frac{2\cot\theta_W}{S-M_Z^2}(v-2a\lambda)$ &
$-\beta_W\frac{\sqrt{S}}{M_W}$ &
$-\beta_W\frac{\sqrt S}{M_W}$ \\
\hline
$-\frac{x_\gamma+y_\gamma}{S}+
\frac{x_Z+y_Z+2\delta_Z}{S-M_Z^2}(v-2a\lambda)$ &
$-\beta_W\frac{\sqrt S}{M_W}$ &
$-\beta_W\frac{\sqrt S}{M_W}$ \\
\hline
\end{tabular}
\label{tab:tab1}
\end{table}
Integrating over the angle $\phi_W$, and assuming $P_L^\prime=0$, the
differential cross section reads
\begin{equation}
\frac{d\sigma}{d\cos\theta}=\frac{1}{4}\left[\left(1+P_L\right)
\frac{d\sigma^+}{d\cos\theta}+\left(1-P_L\right)
\frac{d\sigma^-}{d\cos\theta}\right]\label{longi}\end{equation}
where
\begin{equation}
\frac{d\sigma^{+,-}}{d\cos\theta}=\frac{\vert\vec p\vert}{4\pi s\sqrt s}
\vert {\cal A}^{+,-}\vert^2.\label{pm}\end{equation}
In practice, the initial electron longitudinal polarization
$P_L$ will not be exactly equal to unity, so that the measured cross section
will be a linear combination of $\sigma^+$ and $\sigma^-$ as in
Eq.~(\ref{longi}), with $\vert P_L\vert<1$. In what follows, we shall refer
to `right-handed' ($\sigma^R$) and `left-handed' ($\sigma^L$) cross sections
the cases $P_L=0.9$ and $P_L=-0.9$, respectively. Such values of $P_L$
seem to be obtainable at the NLC \cite{prescott}.\par
Concerning the possibility of exploiting transverse beam polarization,
which will be taken into account for LEP~200 only, a suitable observable is
the azimuthal asymmetry $A_T$, defined as
\begin{equation}
\frac{d(\sigma A_T)}{d\cos\theta}=2\int_0^{2\pi}\frac{d^2\sigma}{d\cos\theta
d\phi_W}\cos(2\phi_W)d\phi_W=P_TP_T^\prime\frac{\vert\vec p\vert}
{4\pi s\sqrt s}{\rm Re}\hskip 2pt ({\cal A}^+{\cal A}^{-*}).
\label{trans}\end{equation}
In our numerical results we shall assume $P_T=P_T^\prime=92.4\%$, which is
the maximum attainable value.

\section{Bounds on anomalous couplings from NLC}
Present constraints on anomalous couplings are obtained by taking only one
or two of them at a time as independent free parameters, and fixing the
remaining ones at the SM values or, alternatively, by assuming specific
models where the couplings are related to each other so that the number of
degrees of freedom is reduced. Bounds derived in this way, although
seemingly stringent, might not fully represent the real situation that can
occur in general. Indeed, when allowing for more than one anomalous coupling,
correlations among these parameters and/or accidental cancellations can
possibly reduce the sensitivity, if a restricted set
of observables, like the unpolarized differential or the total cross section,
is considered. To the purpose of making a significant test by disentangling the
various couplings, it should be desirable to apply a model independent
analysis, where all trilinear gauge boson couplings of Eq.~(\ref{lagra}) are
included and allowed to vary independently. In this regard, as we shall see
below, polarization not only allows to disentangle the bounds for the
different constants in a simple, analytic, way, but also leads to definite
improvements in the accuracy of the constraints. \par
Using Tab.~\ref{tab:tab1} one easily finds that, for specific initial and
final states polarizations, the deviations from the SM of the $\gamma$ and
$Z$ exchange amplitudes depend on the following combinations of anomalous
couplings:
\begin{equation}\ \ \Delta{\cal A}_{LL}^{a}(\gamma)\propto x_{\gamma}
\hskip 2pt ;
\qquad\ \ \ \ \ \
\Delta{\cal A}_{LL}^{a}(Z)\propto\left(x_Z+\delta_Z\hskip 1pt
\frac{3-\beta_W^2}{2}\right)\hskip 1pt g_e^a\label{deltall}\end{equation}
\begin{equation}\Delta{\cal A}_{TL}^{a}(\gamma)\propto x_{\gamma}+y_{\gamma}
\hskip 2pt;\qquad
\Delta{\cal A}_{TL}^{a}(Z)\propto\left(x_Z+y_Z+2\hskip 1pt\delta_Z\right)
\hskip 1pt g_e^a\label{deltatl}\end{equation}
\begin{equation}\ \ \Delta{\cal A}_{TT}^{a}(\gamma) \propto y_{\gamma}
\hskip 2pt ; \qquad\ \ \ \ \ \
\Delta{\cal A}_{TT}^{a}(Z) \propto \left(y_Z+\delta_Z\hskip 1pt
\frac{1-\beta_W^2}{2}\right)\hskip 1pt g_e^a.\label{deltatt}\end{equation}
In Eqs.~(\ref{deltall})-(\ref{deltatt}) the lower indices $LL$, $TL$ and $TT$
refer to the final $W^-W^+$ polarizations, and the upper index $a$ indicates
the initial $e^-$ right-handed ($^+$) or left-handed ($^-$) polarizations,
with $g_e^R=s_W/c_W$ and $g_e^L=g_e^R\left(1-1/2s_W^2\right)$ the
corresponding electron couplings to the $Z$. \par
In order to assess the sensitivity of the different cross sections to the
gauge boson couplings we divide the experimentally significant range of the
production angle $\cos\theta$
(which we take as $\vert\cos\theta\vert\leq 0.98$) into `bins',
and define the $\chi^2$ function:
\begin{equation}
\chi^{2}=\sum^{bins}_i\left[\frac{N_{SM}(i)-N_{anom}(i)}
{\delta N_{SM}(i)}\right]^2.\label{chi2}\end{equation}
As it is conventional in this kind of analyses, the range of $\cos\theta$ is
divided into 10 equal bins for the NLC and into 6 bins for the LEP~200 case.
In Eq.~(\ref{chi2}), in a self-explaining notation
$N(i)= L_{int}\sigma_i\varepsilon_W$ is the expected number of events in the
$i$-th bin, with $\sigma_i$  the corresponding cross section (either the SM
or the anomalous one):
\begin{equation}
\sigma_i\equiv\sigma(z_i,z_{i+1})=
\int \limits_{z_i}^{z_{i+1}}\left({d\sigma}\over{dz}\right)dz,
\label{sigmai}
\end{equation}
where $z=\cos\theta$. For convenience, in the Appendix we give the explicit
expressions for the polarized integrated cross sections $\sigma(z_i,z_{i+1})$
with nonzero anomalous gauge boson couplings. The parameter $\varepsilon_W$
introduced above is the efficiency for $W^+W^-$ reconstruction in the
considered polarization state. We take the channel of
lepton pairs ($e\nu+\mu\nu$) plus two hadronic jets, and correspondingly a
reference value $\varepsilon_W\simeq 0.3$ \cite{frank},
\cite{forty}-\cite{anlauf}, as obtained from the relevant branching ratios.
The actual value of $\varepsilon_W$ for polarized
final states might be considerably smaller, depending on experimental details
\cite{frank}, but definite estimates are presently not available. As a
compensation, for the luminosity $L_{int}$, which everywhere appears
multiplied by $\varepsilon_W$, we make the rather conservative choice compared
with recent findings \cite{ruth}:
\begin{eqnarray}
\int_{}^{}Ldt= 20\ fb^{-1}~(NLC~ 500 ),\nonumber \\
\int_{}^{}Ldt= 50\ fb^{-1}~(NLC~ 1000 ).
\label{lum}
\end{eqnarray}
Finally, in Eq.~(\ref{chi2}), the uncertainty on the number of events
$\delta N_{SM}(i)$ combines both statistical and systematic errors for the
$i$-th bin:
\begin{equation} \delta N_{SM}(i)=
\sqrt{N_{SM}(i)+\left(\delta_{syst}N_{SM}(i)\right)^2},
\label{deltan} \end{equation}
and the systematic error will be taken as $\delta_{syst}=2\%$.\par
As a criterion to derive allowed regions for the coupling constants, we will
impose that $\chi^2\leq\chi^2_{crit}$, where $\chi^2_{crit}$ is a number
that specifies a chosen confidence level and in principle can depend on the
kind of analysis. Eqs.~(\ref{deltall})-(\ref{deltatt}) show that each polarized
cross section involves two well-defined combinations of anomalous couplings
at a time, namely ($x_{\gamma},x_Z+\delta_Z(3-\beta_W^2)/2$),
($x_{\gamma}+y_{\gamma},x_Z+y_Z+2\delta_Z$) and
($y_{\gamma},y_Z+\delta_Z(1-\beta_W^2)/2$). Correspondingly, with two
independent degrees of freedom, in each separate case bounds at the 95\% CL are
obtained by choosing $\chi^2_{crit}=6$ \cite{pdg,cuypers}.\footnote{This
should be compared with the case of only one free parameter, which occurs in
various models, where $\chi^2_{crit}=4$ should be taken to obtain the bounds
at the same CL.} The same $\chi^2_{crit}=6$ is taken in order to derive 95\% CL
bounds on the coupling constants from the combination of both initial
longitudinal polarizations, $d\sigma^R/dz$ ($P_L=0.9$) and
$d\sigma^L/dz$ ($P_L=- 0.9$), for which the combined $\chi^2$ function is
defined as the sum $\chi^2=\chi^2_R+\chi^2_L$.\par
We start the presentation of our numerical results from the case of
longitudinally polarized $W$'s, $e^-e^+\to W^-_LW^+_L$, for both
possibilities of electron beam longitudinal polarization. The resulting area
allowed to the combinations of anomalous couplings in Eq.~(\ref{deltall}) at
the $95\%$ CL is depicted in Fig.~2, for both $\sqrt s=0.5$ and
$1\hskip 2pt TeV$. Actually, as discussed in Ref.~\cite{pankov3}, one
finds elliptical contours which would give four common intersections as
allowed regions, three of them not containing the SM values. Obviously, we
are concentrating here on the region surrounding zero values for anomalous
couplings. This information is not yet sufficient to disentangle
the individual couplings, since from Fig.~2 we simply find the pair of
inequalities
\begin{equation}-\alpha_1^{LL}<x_{\gamma}<\alpha_2^{LL},\label{ll1}
\end{equation}
\begin{equation}-\beta_1^{LL}<x_Z+\delta_Z\frac{3-\beta_W^2}{2}<\beta_2^{LL},
\label{ll2}\end{equation}
so that only $x_\gamma$ is separately constrained at this stage.
Here, $\alpha_{1,2}^{LL}$ and $\beta_{1,2}^{LL}$ are the projections of the
combined allowed area on the horizontal and vertical axes, respectively,
and their values can be directly read from Fig.~2.\par
Turning to the other polarized cross sections, we repeat the same analysis
there. From $e^+e^-\to W^+_TW^-_L+W^+_LW^-_T$ we obtain the allowed region
for the combinations of coupling constants in Eq.~(\ref{deltatl}), depicted in
Fig.~3. This leads to the following inequalities, analogous to
Eqs.~(\ref{ll1}) and (\ref{ll2}):
\begin{equation}-\alpha_1^{TL}<x_{\gamma}+y_{\gamma}<\alpha_2^{TL},
\label{tl1}\end{equation}
\begin{equation}
-\beta_1^{TL}<x_Z+y_Z+2\delta_Z<\beta_2^{TL}.\label{tl2}\end{equation}
\par Finally, from $e^+e^-\to W^+_TW^-_T$ one obtains for the combinations of
coupling constants in Eq.~(\ref{deltatt}) the allowed regions depicted in
Fig.~4, and the corresponding inequalities:
\begin{equation}-\alpha_1^{TT}<y_{\gamma}<\alpha_2^{TT},\label{tt1}
\end{equation}
\begin{equation}
-\beta_1^{TT}<y_Z+\frac{1-\beta_W^2}{2}\delta_Z<\beta_2^{TT}.
\label{tt2}\end{equation}
One can notice that, with initial state polarization, the channel
$e^+e^-\to W^+_TW^-_T$ can separately constrain $y_{\gamma}$.
The limits in Fig.~4 are less restrictive compared to the previous cases,
because they are determined by the larger width of the region allowed by
the left-handed cross section, which is dominated by the
$\vert\tau-\tau^\prime\vert=2$ amplitude ${\cal A}_2(\nu)$
(see Eq.~(\ref{sig})) and therefore has a reduced sensitivity to anomalous
couplings. Moreover, comparing Fig.~4 to Figs.~2 and 3, one can notice that
the bound resulting from $\sigma^R$ has now a quite different shape. This is
the dramatic effect of the contamination of the right-handed cross section
by the much bigger left-handed one for $P_L$ not exactly equal to unity
($P_L=0.9$), as it can be seen from Eq.~(\ref{longi}). \par
By combining Eqs.~(\ref{ll2})-(\ref{tt2}), one can very simply disentangle
the bounds for $\delta_Z$, $x_Z$ and $y_Z$:
\begin{equation}-\frac{1}{\beta_W^2}B_2<\delta_Z<\frac{1}{\beta_W^2}B_1,
\label{B1}\end{equation}
\begin{equation}-\left(\beta_1^{LL}+\frac{3-\beta_W^2}{2\beta_W^2}B_1\right)<
x_Z<\beta_2^{LL}
+\frac{3-\beta_W^2}{2\beta_W^2}B_2\label{B2},\end{equation}
\begin{equation}-\left(\beta_1^{TT}+\frac{1-\beta_W^2}{2\beta_W^2}B_1\right)
<y_Z<\beta_2^{TT}
+\frac{1-\beta_W^2}{2\beta_W^2}B_2,\label{B3}\end{equation}
where $B_1=\beta_1^{LL}+\beta_1^{TT}+\beta_2^{TL}$ and
$B_2=\beta_2^{LL}+\beta_2^{TT}+\beta_1^{TL}$. Adding these constraints to
those in Eqs.~(\ref{ll1}) and (\ref{tt1}) for $x_\gamma$ and $y_\gamma$,
we finally obtain separate bounds for the five anomalous couplings that
determine the general expansion of Eq.~(\ref{lagra}). In this regard, we
should notice the simplicity of this procedure to determine separate
constraints on the trilinear couplings. \par
Actually, in addition to Eq.~(\ref{tt1}) there is one more condition on
$y_\gamma$ from the combination of Eqs.~(\ref{ll1}) and (\ref{tl1}):
\begin{equation}-\left(\alpha_1^{TL}+\alpha_2^{LL}\right)<y_\gamma<
\alpha_1^{LL}+\alpha_2^{TL}.\label{ygamma}\end{equation}
Numerically, which of the two is the most restrictive one depends on the
value of the center of mass energy: indeed, it turns out that for
$\sqrt s=500\hskip 2pt GeV$ the most stringent bound on $y_\gamma$ is
determined by Eq.~(\ref{ygamma}), while Eq.~(\ref{tt1}) gives the most
restrictive condition for $1\hskip 2pt TeV$.\par
The numerical results from these relations, and the chosen inputs for the
luminosity and the initial polarization quoted previously, are summarized in
Tab.~\ref{tab:tab2}.
\begin{table}
\centering
\caption{ Model independent limits on the five $CP$ even nonstandard
gauge boson couplings at the $95\%$ CL.}
\begin{tabular}{|c|c|c|c|c|c|}
\hline
$\sqrt{s}\ (TeV)$ & $x_\gamma\hskip2pt (10^{-3})$ & $y_\gamma\hskip2pt
(10^{-3})$ & $\delta_Z\hskip2pt (10^{-3})$ & $x_Z\hskip2pt (10^{-3})$ &
$y_Z\hskip2pt (10^{-3})$
\\ \hline
$0.5$ & $-2.0\div 2.2$ & $-11.0\div 10.6$ &$-52\div 45$
& $-51\div 59$ & $-22\div 30$
\\ \hline
$1$ &$-0.6\div 0.6$ & $-3.2\div 3.4$ &$-19\div 16$
& $-18\div 20$ & $-5.7\div 6.2$
\\ \hline
\end{tabular}
\label{tab:tab2}
\end{table}
\par In a previous, model independent, analysis of $CP$ conserving anomalous
couplings \cite{pankov3}, instead of the binning procedure followed here we
used polarized cross sections integrated in angular ranges appropriately
chosen in order to optimize the sensitivity to these parameters. Numerically,
the results are qualitatively comparable, but the binning procedure leads
to constraints improved by 10-50\%, depending on the particular
case.\footnote{The possibility to derive a separate bound by a similar
analysis also on the anapole coupling $z_Z$ (in the notation of
Ref. \cite{gounaris1}), which violates both $C$ and $P$
but conserves $CP$, was previously considered in Ref.~\cite{pankov3}. $CP$ odd
anomalous $WW\gamma$ couplings are independently (and stringently) constrained
by the limit on the neutron electric dipole moment \cite{boudjema}.}
\par It should be interesting to specialize the
procedure outlined above to the discussion of few model examples for
nonstandard anomalous trilinear gauge boson couplings, where the number of
such parameters is decreased. A popular framework is that in which anomalous
values of the couplings reflect some new interaction effective at a mass scale
$\Lambda$ much higher than the Fermi scale. Correspondingly, at our (lower)
energy scales, such effects represent corrections to the SM suppressed by
inverse powers of $\Lambda$. As a natural requirement, given the observed
phenomenological success of $SU(2)\times U(1)$, such a gauge symmetry
(spontaneously broken and with $\gamma$, $W$ and $Z$ the corresponding gauge
bosons) is imposed also on the new interactions
\cite{zeppenfeld1}.\footnote{Alternatives to imposing this symmetry have also
been considered, see, e.g., Ref.~\cite{frere}.}
The weak interaction is then described by an effective Lagrangian of the form:
\begin{equation}{\cal L}_W={\cal L}_{SM}+\sum_d\sum_k
\frac{f^{(d)}_k}{\Lambda^{d-4}}{\cal O}^{(d)}_k, \label{lagra1}\end{equation}
where ${\cal L}_{SM}$ is the SM interaction and the second term is the source
of anomalous trilinear gauge boson couplings. This term takes the
form of an expansion in inverse powers of $\Lambda$, where ${\cal O}^{(d)}_k$
are dimension $d$ gauge invariant operators made of $\gamma$,
$W$, $Z$ and Higgs fields, and $f^{(d)}_k$ are coupling constants, not fixed
by the symmetry. From the good agreement of the measured lepton couplings
with the SM ones, one assumes that such couplings remain unaffected by the
new physics. Truncation of the sum in Eq.~(\ref{lagra1}) to the lowest
significant dimension, $d=6$, limits the number of allowed independent
operators (and their corresponding constants) to three
\cite{zeppenfeld1,ruj,buch,gounaris3}:
\begin{eqnarray}{\cal O}^{(6)}_{WWW}&=&Tr\left[{\hat W}_{\mu\nu}
{\hat W}^{\nu\rho}{\hat W}^{\mu}_{\rho}\right],\nonumber\\
{\cal O}^{(6)}_W&=&\left(D_\mu\Phi\right)^\dagger{\hat W}^{\mu\nu}
\left(D_\nu\Phi\right),\nonumber\\
{\cal O}^{(6)}_B&=&\left(D_\mu\Phi\right)^\dagger{\hat B}^{\mu\nu}
\left(D_\nu\Phi\right).\label{oper}\end{eqnarray}
Here, $\Phi$ is the Higgs doublet and, in terms of the $B$ and $W$
field strengths: ${\hat B}^{\mu\nu}=i(g^\prime/2)B^{\mu\nu}$,
${\hat W}^{\mu\nu}=i(g/2){\vec\tau}\cdot{\vec W}^{\mu\nu}$ with $\vec\tau$
the Pauli matrices. Therefore, in such a model, only three
anomalous couplings are independent:
\begin{equation}x_\gamma=\cos^2\theta_W\hskip 2pt
\left(f^{(6)}_B+f^{(6)}_W\right)\hskip 2pt\frac{M^2_Z}{2\Lambda^2};
\qquad y_\gamma=f^{(6)}_{WWW}
\hskip 2pt\frac{3M^2_Wg^2}{2\Lambda^2}; \label{deltaz}\end{equation}
\begin{equation}\delta_Z=\cot\theta_W\hskip 2pt f^{(6)}_W
\hskip 2pt\frac{M^2_Z}{2\Lambda^2};
\quad x_Z=-\tan\theta_W\hskip 2pt x_\gamma;
\quad y_Z=\cot\theta_W\hskip 2pt y_\gamma.\label{xgamma}\end{equation}
According to Eqs.~(\ref{deltaz}) and (\ref{xgamma}), in this model there are
only three independent couplings which can be chosen to be
$x_\gamma$, $y_\gamma$ and $\delta_Z$. As mentioned in Ref.~\cite{zeppenfeld1},
the correlations between different anomalous trilinear gauge boson couplings
exhibited in Eqs.~(\ref{deltaz}) and (\ref{xgamma}) are due to the truncation
of the effective Lagrangian (\ref{lagra1}) at the dimension 6 level, and do not
hold any longer when dimension 8 (or higher) operators are included.\par
Further reduction in the number of the anomalous couplings occurs in the
`HISZ scenario' \cite{zeppenfeld1}, where the relation $f^{(6)}_B=f^{(6)}_W$
in Eqs.~(\ref{deltaz}) and (\ref{xgamma}) is assumed. In this case, the
$WWZ$ couplings are so related:
\begin{equation}
\delta_Z=\frac{1}{2\sin\theta_W\cos\theta_W}\hskip 2pt x_\gamma,
\quad x_Z=-\tan\theta_W\hskip 2pt x_\gamma,
\quad y_Z=\cot\theta_W\hskip 2pt y_\gamma.
\label{hisz}\end{equation}
\par Another way to reduce the number of independent trilinear anomalous
couplings starts from imposing just global $SU(2)_L$ symmetry on the
Lagrangian in Eq.~(\ref{lagra}). This directly implies the relation
$x_Z=-\tan\theta_W\hskip 2pt x_\gamma$, the same as in Eq.~(\ref{xgamma}).
Further reduction is obtained by neglecting dimension 6 quadrupole
operators, so that $y_\gamma=y_Z=0$, and by cancelling the order $s^2$
tree-level unitarity violating contributions to $WW$ scattering, which in turn
leads to the condition
$\delta_Z=x_\gamma/\sin\theta_W\hskip 1pt\cos\theta_W$
\cite{kuroda,bilchak}.\par
For the model with three parameters, the region allowed to
$(x_\gamma,\delta_Z)$, presented in Fig.~5, corresponds to $W_LW_L$
production, combining both left-handed and right-handed initial polarization.
Comparing to the results in Tab.~\ref{tab:tab2}, we notice that $\delta_Z$
can be more tightly constrained in this case than in the general one.
Concerning the third independent coupling, $y_\gamma$, the best bounds are
obtained from the combination of $W_LW_L$ and $W_LW_T$ production channels.
In the case of the two-parameter model of Ref.~\cite{kuroda}, the bounds on
$x_\gamma$ and $\delta_Z$ are obtained in the same way as above, and are
numerically identical.\par
The bounds relevant to the two-parameter model of Ref.~\cite{zeppenfeld1} are
shown in Fig.~6. In this case, due the relation (\ref{hisz}) among
the couplings, $\sigma^L$ numerically proves to be more sensitive than
$\sigma^R$. Concerning final state polarizations, the bound on $x_\gamma$ is
obtained from $W_LW_L$ production, while that on $y_\gamma$ involves the
combination of both $LL$ and $TL+LT$ polarized cross sections. For an
illustration, in Fig.~6 we also report the region allowed by the cross
section for unpolarized $W$'s.\par
Tab.~\ref{tab:tab3} summarizes the numerical bounds that can be obtained
from our analysis for the models of anomalous couplings considered here.
\begin{table}
\centering
\caption{Limits on anomalous gauge boson couplings at the 95\% CL
for the models with three, two and one independent parameters.}
\begin{tabular}{|c|c|c|c|c|c|}
\hline
\multicolumn{6}{|c|}{Model with three
independent anomalous constants \cite{zeppenfeld1}:
$x_\gamma$, $y_\gamma$, $\delta_Z$;}\\
\multicolumn{6}{|c|}{
$x_Z=-\tan\theta_W\hskip 2pt x_\gamma$, $y_Z=\cot\theta_W\hskip 2pt
y_\gamma$.}\\ \hline
$\sqrt{s}\ (TeV)$ & $x_\gamma\hskip 2pt (10^{-3})$ & $\delta_Z\hskip 2pt
(10^{-3})$& $x_Z\hskip 2pt (10^{-3})$
& $y_\gamma\hskip 2pt (10^{-3})$& $y_Z\hskip 2pt (10^{-3})$
\\ \hline
$0.5$ & $-2.0\div 2.2$ & $-3.8\div 3.8$ & $-1.2\div 1.1$
&$-7.0\div 7.5$& $-12.8\div 13.7$
\\ \hline
$1  $ & $-0.6\div 0.6$ & $-1.1\div 1.1$ & $-0.3\div 0.3$
&$-4.0\div 4.5$ & $-7.3\div 8.2$
\\ \hline
\hline
\multicolumn{6}{|c|} {Model with
two independent anomalous constants \cite{zeppenfeld1}:  $x_\gamma$,
$y_\gamma$;}\\
\multicolumn{6}{|c|}{$\delta_Z=x_\gamma/2\sin\theta_W\cos\theta_W$,
$x_Z=-\tan\theta_W\hskip 2pt x_\gamma$, $y_Z=\cot\theta_W\hskip 2pt y_\gamma$.}
 \\ \hline
$\sqrt{s}\ (TeV)$ & $x_\gamma\hskip 2pt
(10^{-3})$ & $\delta_Z\hskip 2pt (10^{-3})$ & $x_Z\hskip 2pt(10^{-3})$
& $y_\gamma\hskip 2pt (10^{-3})$ & $y_Z\hskip 2pt (10^{-3})$
\\ \hline
$0.5$ & $-1.8\div 1.8$ & $ -2.1\div 2.1 $ & $ -1.0\div 1.0$&$-6.6\div 6.8$&
$-12.1\div 12.4$    \\ \hline
$1$ & $-0.5\div 0.5$ & $ -0.6\div 0.6$ &$ -0.3\div 0.3$&$-3.0\div 2.4$&
$-5.5\div 4.4$\\ \hline
\hline
\multicolumn{6}{|c|}{
Model with one independent anomalous constant \cite{bilchak}:
$x_\gamma;$}\\
\multicolumn{6}{|c|}{
$x_Z=-\tan\theta_W$\hskip 2pt $x_\gamma=-\sin^2\theta_W$\hskip 2pt
$\delta_Z$.}\\
\hline
$\sqrt{s}\ (TeV)$ & $x_\gamma \hskip 2pt
(10^{-3})$ & $\delta_Z\hskip 2pt (10^{-3})$ & $x_Z\hskip 2pt(10^{-3})$
& $y_\gamma\hskip 2pt$ & $y_Z\hskip 2pt $
\\ \hline
$0.5$ & $-1.1\div 1.1$ & $  -2.6\div 2.6 $ &$ -0.6\div 0.6 $
& 0&0 \\ \hline
$1$ & $-0.3\div 0.3$ & $ -0.8 \div 0.8 $ &$-0.2  \div 0.2 $
&0&0 \\ \hline
\end{tabular}
\label{tab:tab3}
\end{table}

\section{Bounds on anomalous couplings from LEP~200}
At this facility, no initial beam longitudinal polarization is planned
\cite{treille}. As anticipated in Sect.~1, to perform a model independent
analysis of all five $CP$-even couplings following the procedure above,
$\sigma^{unpol}$ can play the role of $\sigma^L$ (having a similar dependence
on these couplings), and the azimuthal asymmetry $A_T$ in Eq.~(\ref{trans})
can be combined with $\sigma^{unpol}$ to give the bounds. Thus, we
assume that the transverse polarization of initial beams will be
available, and that the final $W$'s polarizations could be measured with the
same efficiency used in the previous Sections. Due to the limited statistics
provided by the luminosity at LEP~200:
\begin{equation} \int_{}^{}Ldt= 500\ pb^{-1}~(LEP~ 200),\label{lumino}
\end{equation}
we take 6 equal bins in order to have a significant number of events per
beam and, furthermore, we assume the same systematic uncertainty as in
Eq.~(\ref{deltan}) as well as the same reconstruction efficiency
$\varepsilon_W$. By performing the same kind analysis presented in the
previous Section, we would find for the combinations of anomalous couplings
relevant to Eqs.~(\ref{deltall})-(\ref{deltatt}) the 95\% CL allowed regions
presented in Figs.~7-9, respectively. These are the analogues of Figs.~2-4
for the case of NLC. Quite similarly, the constraints at LEP correspond to the
combinations of the bounds from $A_T$ and $\sigma^{unpol}$ for $W_LW_L$,
$W_LW_T+W_TW_L$ and $W_TW_T$ production, respectively. In the last case, from
Fig.~9 one can notice that the azimuthal asymmetry is not so helpful to
minimize the combined allowed region, which therefore in almost entirely
determined by $\sigma^{unpol}$.\par
By combining the analogues of Eqs.~(\ref{ll1})-(\ref{ygamma}), one can
disentangle the bounds for the different couplings constants. The numerical
results are presented in Tab.~\ref{tab:tab4} for two values of the CM
energy, namely $\sqrt s=200\hskip 2pt GeV$ and $230\hskip 2pt GeV$, and the
luminosity in Eq~(\ref{lumino}). As expected, the constraints become more
stringent with increasing energy.\par
\begin{table}
\centering
\caption{ Model independent limits on the five $CP$ even nonstandard
gauge boson couplings at the $95\%$ CL for LEP~200.}
\begin{tabular}{|c|c|c|c|c|c|}
\hline
$\sqrt{s}\ (GeV)$ & $x_\gamma\hskip2pt (10^{-1})$ & $y_\gamma\hskip2pt
(10^{-1})$ & $\delta_Z\hskip2pt (10^{-1})$ & $x_Z\hskip2pt (10^{-1})$ &
$y_Z\hskip2pt (10^{-1})$
\\ \hline
$200$ & $-0.9\div 1.0$ & $-2.0\div 2.9$ &$-26.4\div 23.6$
& $-32.8\div 36.7$ & $-10.6\div 13.1$
\\ \hline
$230$ &$-0.5\div 0.6$ & $-1.4\div 2.0$ &$-15.6\div 13.8$
& $-18.5\div 20.8$ & $-5.7\div 7.6$
\\ \hline
\end{tabular}
\label{tab:tab4}
\end{table}
\par
Concerning the application of this approach to models with a reduced number
of independent anomalous couplings, the expected sensitivities, for the same
model examples considered in the previous Section, are exposed in
Tab.~\ref{tab:tab5}.
\begin{table}
\centering
\caption{Limits on anomalous gauge boson couplings at the 95\% CL
for the models with three, two and one independent parameters for LEP~200.}
\begin{tabular}{|c|c|c|c|c|c|}
\hline
\multicolumn{6}{|c|}{Model with three
independent anomalous constants \cite{zeppenfeld1}:
$x_\gamma$, $y_\gamma$, $\delta_Z$;}\\
\multicolumn{6}{|c|}{
$x_Z=-\tan\theta_W\hskip 2pt x_\gamma$, $y_Z=\cot\theta_W\hskip 2pt
y_\gamma$.}\\ \hline
$\sqrt{s}\ (GeV)$ & $x_\gamma\hskip 2pt
(10^{-1})$ & $\delta_Z\hskip 2pt (10^{-1})$ & $x_Z\hskip 2pt(10^{-1})$
& $y_\gamma\hskip 2pt (10^{-1})$ & $y_Z\hskip 2pt(10^{-1}) $
\\ \hline
$200$ & $-0.86\div 0.94$ & $ -1.2\div 1.3 $ & $ -0.51\div 0.47$&$-1.4\div 2.2$&
$-2.56\div 4.02$    \\ \hline
$230$ & $-0.52\div 0.62$ & $ -1.0\div 1.1$ &$ -0.34\div 0.28$&$-1.0\div 2.0$&
$-1.8\div 3.66$\\ \hline
\hline
\multicolumn{6}{|c|} {Model with
two independent anomalous constants \cite{zeppenfeld1}:  $x_\gamma$,
$y_\gamma$;}\\
\multicolumn{6}{|c|}{$\delta_Z=x_\gamma/2\sin\theta_W\cos\theta_W$,
$x_Z=-\tan\theta_W\hskip 2pt x_\gamma$, $y_Z=\cot\theta_W\hskip 2pt y_\gamma$.}
 \\ \hline
$\sqrt{s}\ (GeV)$ & $x_\gamma\hskip 2pt
(10^{-1})$ & $\delta_Z\hskip 2pt (10^{-1})$ & $x_Z\hskip 2pt(10^{-1})$
& $y_\gamma\hskip 2pt (10^{-1})$ & $y_Z\hskip 2pt (10^{-1}) $
\\ \hline
$200$ & $-0.6\div 0.7$ & $ -0.71\div 0.83 $ & $ -0.38\div 0.33$&$-1.5\div 1.5$&
$-2.7\div 2.2$    \\ \hline
$230$ & $-0.42\div 0.48$ & $ -0.5\div 0.57$ &$ -0.26\div 0.23$&$-1.1\div 1.2$&
$-2.0\div 2.2$\\ \hline
\hline
\multicolumn{6}{|c|}{
Model with one independent anomalous constant \cite{bilchak}:
$x_\gamma;$}\\
\multicolumn{6}{|c|}{
$x_Z=-\tan\theta_W$\hskip 2pt $x_\gamma=-\sin^2\theta_W$\hskip 2pt
$\delta_Z$.}\\
\hline
$\sqrt{s}\ (GeV)$ & $x_\gamma \hskip 2pt
(10^{-1})$ & $\delta_Z\hskip 2pt (10^{-1})$ & $x_Z\hskip 2pt(10^{-1})$
& $y_\gamma\hskip 2pt$ & $y_Z\hskip 2pt $
\\ \hline
$200$ & $-0.39\div 0.41$ & $  -0.93\div 0.97 $ &$ -0.22\div 0.21 $
& 0&0 \\ \hline
$230$ & $-0.30\div 0.33$ & $ -0.71 \div 0.78 $ &$-0.18  \div 0.16 $
&0&0 \\ \hline
\end{tabular}
\label{tab:tab5}
\end{table}

\section{Concluding remarks}
One of the basic points of the analysis presented above is the use of
final $W^\pm$ polarization to group the five independent anomalous
trilinear gauge boson couplings into pairs of `effective' combinations as in
Eqs.~(\ref{deltall})-(\ref{deltatt}), {\it via} the specific dependence of
the helicity amplitudes relevant to the considered differential cross
sections. Such cross sections for polarized $W$'s should be obtained
experimentally from angular distributions of the $W^\pm$ decay products
\cite{bilenky1}. This leads to a simplified two-dimensional analysis
(rather than a three- or a five-dimensional one) for each final polarization
and, by a $\chi^2$ procedure, bounds in the two-parameter planes of the
corresponding pairs of `effective' coupling constants are obtained.\par
The initial electron beam polarization (either longitudinal or transverse)
turns out to have a fundamental role in drastically reducing the above
mentioned two-dimensional allowed regions. Finally, by combining
Eqs.~(\ref{ll1}) to (\ref{tt2}), one can obtain separate bounds for each
of the five $CP$ even couplings. Thus in summary, while the specific dependence
of the final state polarization on the anomalous couplings allows a model
independent analysis of the general case, initial beam polarization can
be used to further restrict the bounds. \par
{}From the numerical point of view, the bounds presented in Tabs.~2 to 5 are
rather stringent and clearly, for a more complete test of the SM, the
electroweak corrections \cite{fleischer} can be included in the analysis.
Furthermore, the sensitivity to anomalous couplings indicated by these results
crucially depends on the chosen inputs, in particular on the assumed value of
the polarized $W^\pm$ reconstruction efficiency, so that the analysis
needs to be supplemented by a more detailed knowledge of the experimental
performances. \par
Finally, we recall that the procedure presented here is based on the
differential $W^+W^-$ production cross section. However, looking for
further increased sensitivity to the anomalous couplings, it  might be
worthwhile to apply a similar analysis to more detailed observables including
angular distributions of $W^+$ and $W^-$ decay products, such as those
considered in Refs.~\cite{bilenky1} and \cite{sekulin}, and try to assess
there the distinguished role of initial $e^+e^-$ polarization.

\begin{center}{\large\bf Acknowledgements}\end{center}
\noindent
Two of us (VVA and AAP) acknowledge the support and the hospitality of
INFN-Sezione di Trieste and of the International Centre for Theoretical
Physics, Trieste. They also acknowledge the support from the International
Science Foundation and Belarussian Government, Grant N. F9D100.

\newpage
\section*{Appendix}

The integrated cross section of process (\ref{proc}) defined in
Eq.~(\ref{sigmai}) can be generally expressed, for arbitrary degrees of
longitudinal polarization of electrons ($P_L$) and positrons ($\tilde P_L$),
as ($z\equiv\cos\theta$):
$$
\sigma(z_1, z_2)=\frac{1}{4}\left[(1+P_L)\cdot
(1-\tilde P_L)\hskip 2pt \sigma^{+}(z_1, z_2) +(1-P_L)\cdot
(1+\tilde P_L)\hskip 2pt \sigma^{-}(z_1, z_2)
\right].\eqno(A1)\label{pcr1}
$$
The corresponding integrated cross sections for polarized final $W$'s,
to be inserted in Eq.~(A1), can be written as follows:
$$
\sigma^{+,-}_{\alpha\beta}(z_1, z_2)=
C\cdot \sum_{i=0}^{i=11} F_i^{+,-}{\cal O}_{i,\ \alpha\beta}(z_1, z_2),
\eqno(A2)\label{pcr2}
$$
where $C=\pi\alpha^2_{em}\beta_W/2s$, the helicities of the initial
$e^+e^-$ and final $W^+W^-$ states are labeled as $+,-$
($\lambda=-\lambda^\prime=\pm 1/2$) and $\alpha\beta=(LL,\ ,TT,\ TL)$,
respectively. In Eq.~(A2) we use the following notation:
${\cal O}_{i,\alpha\beta}(z_1, z_2)\equiv{\cal O}_{i,\alpha\beta}(z_2)-
{\cal O}_{i,\alpha\beta}(z_1)$, where ${\cal O}_{i,\alpha\beta}$
are functions of the kinematical variables which characterize the various
possibilities for the final $W^+W^-$ polarizations (or the sum over all
polarizations for unpolarized $W$'s). The $F_i$ are combinations of coupling
constants, where the anomalous trilinear gauge boson couplings explicitly
appear. For the case of right-handed electrons (and left-handed positrons) we
have, with $\chi_Z$ the $Z$ boson propagator:
$$
F_1^{+} = 2(1-g_Zg^R_e\cdot\chi_Z)^2 \nonumber \\
$$
$$
F_3^{+} = x_{\gamma}-g^R_e(x_Z+x_\gamma g_Z)\cdot\chi_Z+
(g^R_e\cdot\chi_Z)^2 g_Z x_Z \nonumber \\
$$
$$
F_4^{+} = y_{\gamma}-g^R_e(y_Z+y_\gamma g_Z)\cdot\chi_Z+
(g^R_e\cdot\chi_Z)^2 g_Z y_Z \nonumber \\
$$
$$
F_9^{+} = \frac{1}{2}(x_\gamma-g^R_e x_Z\cdot\chi_Z)^2
\nonumber \\
$$
$$
F_{10}^{+} = \frac{1}{2}(y_\gamma-g^R_e y_Z\cdot\chi_Z)^2
$$
$$
  F_{11}^{+} = \frac{1}{2}\left[x_\gamma y_\gamma-
g^R_e(x_\gamma y_Z+x_Z y_\gamma)\cdot\chi_Z+
(g^R_e\cdot\chi_Z)^2 x_Z y_Z\right]\eqno(A3)
$$
The remaining $F^{+}$ are zero.
For the case of left-handed electrons (and right-handed positrons):
$$
F_0^{-} = \frac{1}{16s^4_W} \nonumber \\
$$
$$
F_1^{-} = 2(1-g_Zg^L_e\cdot\chi_Z)^2 \nonumber \\
$$
$$
F_2^{-} = -\frac{1}{2s^2_W}(1-g_Zg^L_e\cdot\chi_Z) \nonumber \\
$$
$$
F_3^{-} = x_{\gamma}-g^L_e(x_Z+x_\gamma g_Z)\cdot\chi_Z+
(g^L_e\cdot\chi_Z)^2 g_Z x_Z \nonumber \\
$$
$$
F_4^{-} = y_{\gamma}-g^L_e(y_Z+y_\gamma g_Z)\cdot\chi_Z+
(g^L_e\cdot\chi_Z)^2 g_Z y_Z \nonumber \\
$$
$$
F_6^{-} = -\frac{1}{4s^2_W}(x_\gamma-x_Z\hskip 2pt g^L_e\cdot\chi_Z)
\nonumber \\
$$
$$
F_7^{-} = -\frac{1}{4s^2_W}(y_\gamma-y_Z\hskip 2pt g^L_e\cdot\chi_Z)
\nonumber \\
$$
$$
F_9^{-} = \frac{1}{2}(x_\gamma-g^L_e x_Z\cdot\chi_Z)^2
\nonumber \\
$$
$$
F_{10}^{-} = \frac{1}{2}(y_\gamma-g^L_e y_Z\cdot\chi_Z)^2
\nonumber \\
$$
$$
F_{11}^{-} = \frac{1}{2}\left[x_\gamma y_\gamma-
g^L_e(x_\gamma y_Z+x_Z y_\gamma)\cdot\chi_Z+
(g^L_e\cdot\chi_Z)^2 x_Z y_Z\right]\eqno(A4)
$$
The remaining $F^{-}$ are zero.
Eqs.~(A3)--(A4) are obtained in the approximation where the imaginary part of
the $Z$ boson propagator is neglected. Accounting for this  effect requires
the replacements $\chi\to Re\chi$ and $\chi^2\to\vert\chi\vert^2$ in
the right-hand sides of Eqs.~(A3)--(A4).\par
In Eq.~(A2), for the longitudinal ($LL$) cross sections
$\sigma(e^+e^-\to W^+_LW^-_L)$:
$$
{\cal O}_{0,LL}(z) = \frac{s}{4M^4_W}\left[s^3(J_0-J_4)-
4M_W^4(3s+4M_W^2)(J_0-J_2)-\right.\nonumber \\
$$
$$
\left.4(s+2M_W^2)\vert\vec p\vert s\sqrt{s}(J_1-J_3)\right] \nonumber \\
$$
$$
{\cal O}_{1,LL}(z) = \frac{s^3-12sM_W^4-16M_W^6}{8sM_W^4}K_1
\nonumber \\
$$
$$
{\cal O}_{2,LL}(z) = \frac{\vert\vec p\vert s
\sqrt {s}(s+2M_W^2)}{2M_W^4}(I_1-I_3)-
\frac{s^3-12sM_W^4-16M_W^6}{4M_W^4}(I_0-I_2)
\nonumber \\
$$
$$
{\cal O}_{3,LL}(z) = \frac{s^2-2M_W^2s-8M_W^4}{2M^4_W}K_1
\nonumber \\
$$
$$
{\cal O}_{4,LL}(z) = {\cal O}_{5,LL}(z)={\cal O}_{7,LL}(z)={\cal O}_{8,LL}(z)=
{\cal O}_{10,LL}(z)={\cal O}_{11,LL}(z)=0 \nonumber \\
$$
$$
{\cal O}_{6,LL}(z) = \frac{s}{2M_W^4}\left[\left(8M_W^4+
2sM_W^2-s^2\right)(I_0-I_2)+
2s\vert\vec p\vert\sqrt{s}(I_1-I_3)\right] \nonumber \\
$$
$$
{\cal O}_{9,LL}(z) = 2\frac{s\vert\vec p\vert^2}{M_W^4}K_1
\eqno(A5)
$$
For the transverse ($TT$) cross sections $\sigma(e^+e^-\to W^+_TW^-_T)$:
$$
{\cal O}_{0,TT}(z) = 4s\left[s(J_0-J_4)-2M_W^2(J_0-J_2)-
2\vert\vec p\vert\sqrt{s}(J_1-J_3)\right] \nonumber \\
$$
$$
{\cal O}_{1,TT}(z) =\frac{M_W^2}{2s}{\cal O}_{4,TT}(z)=
\frac{M_W^4}{s^2}{\cal O}_{10,TT}(z)= \frac{4{\vert\vec p\vert}^2}{s}K_1
\nonumber \\
$$
$$
{\cal O}_{2,TT}(z) =\frac{M_W^2}{s}{\cal O}_{7,TT}(z) =
4\vert\vec p\vert\sqrt{s}(I_1-I_3)
-8{\vert\vec p\vert}^2(I_0-I_2) \nonumber \\
$$
$$
{\cal O}_{3,TT}(z) = {\cal O}_{5,TT}(z)={\cal O}_{6,TT}(z)={\cal O}_{8,TT}(z)=
{\cal O}_{9,TT}(z)={\cal O}_{11,TT}(z)=0 \eqno(A6)
$$
Finally, for the production of one longitudinal plus one transverse
vector boson $(TL+LT)$:
$$
{\cal O}_{0,TL}(z) = \frac{2s}{M^2_W}\left[s^2(J_0+J_4)-
4\vert\vec p\vert\sqrt{s}(4{\vert\vec p\vert}^2 J_1+sJ_3)+\right.
\nonumber \\
$$
$$
\left. 4M_W^4(J_0+J_2)+2s(s-6M_W^2)J_2
-4sM_W^2J_0 \right] \nonumber \\
$$
$$
2{\cal O}_{1,TL}(z) =
{\cal O}_{3,TL}(z) = {\cal O}_{4,TL}(z)={\cal O}_{11,TL}(z)=
2{\cal O}_{9,TL}(z)=2{\cal O}_{10,TL}(z)=\frac{8{\vert\vec p\vert}^2}{M_W^2}K_2
\nonumber \\
$$
$$
{\cal O}_{2,TL}(z) = {\cal O}_{6,TL}(z)={\cal O}_{7,TL}(z)=
\frac{4\vert\vec p\vert\sqrt{s}}{M_W^2}\left[4{\vert\vec p\vert}^2I_1+sI_3-
2{\vert\vec p\vert}\sqrt{s}(I_0+I_2)\right]\nonumber \\
$$
$$
{\cal O}_{5,TL}(z) = \frac{16{\vert\vec p\vert}^3\sqrt{s}z^2}{M_W^4}
\nonumber \\
$$
$$
{\cal O}_{8,TL}(z) = \frac{16s{\vert\vec p\vert}^2}{M_W^4}
\left[M_W^2I_0+2\vert\vec p\vert\sqrt{s}I_1-
(s-M_W^2)I_2\right]\eqno(A7)
$$
In Eqs.~(A5)-(A7) the functions $I$, $J$ and $K$ are
($d=M_W^2-s/2$, $b=s\beta_W^2/2$, $t=d+bz$):
$$
I_0(z)=\frac{1}{b}\log\vert t\vert
\nonumber \\
$$
$$
I_1(z)=\frac{1}{b^2}\left(t-d\log\vert t\vert\right)
\nonumber \\
$$
$$
I_2(z)=\frac{1}{b^3}\left(\frac{t^2}{2}-2dt+d^2\log\vert t\vert\right)
\nonumber \\
$$
$$
I_3(z)=\frac{1}{b^4}
\left(\frac{t^3}{3}-\frac{3dt^2}{2}+3d^2t-d^3\log\vert t\vert\right)
\nonumber \\
$$
$$
J_0(z)=-\frac{1}{bt}
\nonumber \\
$$
$$
J_1(z)=\frac{1}{b^2}\left(\log\vert t\vert+\frac{d}{t}\right)
\nonumber \\
$$
$$
J_2(z)=\frac{1}{b^3}
\left(t-2d\log\vert t\vert-\frac{d^2}{t}\right)
\nonumber \\
$$
$$
J_3(z)=\frac{1}{b^4}
\left(\frac{t^2}{2}-3dt+3d^2\log\vert t\vert+\frac{d^3}{t}\right)
\nonumber \\
$$
$$
J_4(z)=\frac{1}{b^5}
\left(\frac{t^3}{3}-\frac{4dt^2}{2}+6d^2t-4d^3\log\vert t\vert-\frac{d^4}{t}
\right)\nonumber \\
$$
$$
K_{1,2}(z)=z\mp\frac{z^3}{3}
\eqno(A8)
$$

\newpage
\section*{Figure captions}
\begin{description}

\item{\bf Fig.~1} The Feynman diagrams for the $e^+e^-\to W^+W^-$.
\item{\bf Fig.~2} Allowed domains (95\% CL) for
($x_\gamma, x_Z+\delta_Z(3-\beta_W^2)/2$) from $e^+e^-\to W^+_LW^-_L$
with longitudinally polarized electrons at $\sqrt s=0.5\hskip 2pt TeV$ and at
$\sqrt s=1\hskip 2pt TeV$, inputs as specified in the text.
\item{\bf Fig.~3} Allowed domains (95\% CL) for
($x_{\gamma}+y_{\gamma},\hskip 2pt x_Z+y_Z+2\delta_Z$)
from $e^+e^-\to W^+_LW^-_T+W^+_TW^-_L$ with same inputs as in Fig.~2.
\item{\bf Fig.~4} Allowed domains (95\% CL) for
($y_{\gamma},\hskip 2pt y_Z+\delta_Z\hskip 2pt (1-\beta_W^2)/2$)
from $e^+e^-\to W^+_TW^-_T$ with same inputs as in Fig.~2.
\item{\bf Fig.~5} Allowed domains (95\% CL) for ($x_\gamma, \delta_Z$)
for the models with three \cite{zeppenfeld1} and two \cite{kuroda}
independent couplings from $e^+e^-\to W^+_LW^-_L$ with polarized electrons at
$\sqrt s=0.5\hskip 2pt TeV$ and $\sqrt s=1\hskip 2pt TeV$.
\item{\bf Fig.~6} Allowed domains (95\% CL) for ($x_\gamma, y_\gamma$)
for the models with two independent couplings (`HISZ scenario'
\cite{zeppenfeld1}) from $\sigma^L$ of process $e^+e^-\to W^+_LW^-_L$ at
$\sqrt s=0.5\hskip 2pt TeV$. The notation `unpol' refers to unpolarized
$W^\pm$ final states.
\item{\bf Fig.~7}  Allowed domains (95\% CL) for
($x_\gamma, x_Z+\delta_Z(3-\beta_W^2)/2$) from $e^+e^-\to W^+_LW^-_L$
with unpolarized ($\sigma^{unpol}$) and transversely polarized ($A_T$) initial
$e^+e^-$ beams at $\sqrt s=200\hskip 2pt GeV$, inputs as specified in the text.
The hatched allowed area: combination of $\sigma^{unpol}$ and $A_T$.
\item{\bf Fig.~8} Same as Fig.~7, for
($x_{\gamma}+y_{\gamma},\hskip 2pt x_Z+y_Z+2\delta_Z$)
from $e^+e^-\to W^+_LW^-_T+W^+_TW^-_L$.
\item{\bf Fig.~9} Same as Fig.~7, for
($y_{\gamma},\hskip 2pt y_Z+\delta_Z\hskip 2pt (1-\beta_W^2)/2$)
from $e^+e^-\to W^+_TW^-_T$.
\end{description}

\end{document}